\RequirePackage{fix-cm}
\documentclass[smallextended]{svjour3}       
\smartqed  
\usepackage{natbib}
\bibpunct{(}{)}{;}{a}{}{,}
\usepackage{graphicx}
\usepackage{amsmath}
\usepackage{amssymb}
\begin{document}
\title{Centaur and giant planet crossing populations: origin and distribution 
}


\author{Romina P. Di Sisto \and Natalia L. Rossignoli}
\institute{Romina P. Di Sisto \at Facultad de Ciencias Astron\'omicas y Geof\'isicas, Universidad 
Nacional de La Plata \\ Instituto de Astrof\'{\i}sica de La Plata, CCT La Plata-CONICET-UNLP \\
   Paseo del Bosque S/N (1900), La Plata, Argentina. \\
   \email{romina@fcaglp.unlp.edu.ar}\\
   \and   Natalia L. Rossignoli \at
           Instituto de Astrof\'{\i}sica de La Plata, CCT La Plata-CONICET-UNLP \\
   Paseo del Bosque S/N (1900), La Plata, Argentina.}

\date{}

\maketitle
\vspace{-2cm}
\begin{abstract}
The current giant planet region is a transitional zone where
transneptunian objects (TNOs) cross in their way to becoming Jupiter 
Family Comets (JFCs).  Their dynamical behavior is conditioned by the 
intrinsic dynamical features of TNOs and also by the encounters with the
giant planets. We address the Giant Planet Crossing (GPC) population 
(those objects with $5.2$ au $ < q < 30$ au) studying their number and 
their evolution from their sources, considering the current 
configuration of the Solar System. This subject is reviewed from 
previous investigations and also addressed by new numerical simulations 
of the dynamical evolution of Scattered Disk Objects (SDOs). 
We obtain a model of the intrinsic orbital element distribution of GPCs. The Scattered Disk represents the main source of prograde GPCs and Centaurs, while the contribution from Plutinos lies between one and two orders of magnitude below that from the SD. We obtain the number and size distribution of GPCs from our model, computing 9600 GPCs from the SD with $D > 100$ km and $\sim 10^8$ with $D > 1$ km in the current population. The contribution from other sources is considered negligible. 
The mean lifetime in the Centaur zone is 7.2 Myr, while the mean lifetime of SDOs in the GPC zone is of 68 Myr. The latter is dependent on the initial inclination, being the ones with high inclinations the ones that survive the longest in the GPC zone. There is also a correlation of lifetime with perihelion distance, where greater perihelion leads to longer lifetime. The dynamical evolution of observed GPCs is different for prograde and retrograde objects. Retrograde GPCs have lower median lifetime than prograde ones, thus experiencing a comparatively faster evolution. However, it  is  probable  that this faster evolution is due to the fact that the majority of retrograde GPCs have low perihelion values and then, lower lifetimes.


\end{abstract}

\section{Introduction}
\label{intro}

The giant planet region is a vast territory in the Solar System in which it is possible and suitable to test both the Solar System origin and its evolution, as well as specific dynamical and physical processes that take place in the planets, their satellites and in all the minor body populations. In addition, the whole subject is continually fed by new observations and theoretical studies. 

The dynamical evolution of objects in the giant planet region, commonly called ``Centaurs" has been mainly investigated in relation to their ``parental ties", i.e. as progeny of Transneptunian Objects (TNOs) and as progenitors of Jupiter Family Comets (JFCs). Due to their transitional object quality, their dynamical behavior is conditioned by the intrinsic dynamical features of TNOs. Neptune can be considered the ``nexus" between TNOs and Centaurs, since the eventual gravitational interactions of some TNOs with this planet can transfer them to the planetary zone, becoming then Centaur objects. Although the evolutionary path of TNOs from the transneptunian (TN) region into the Centaur region and then into the JFC population is well studied and accepted, the actual number and size distribution of Centaurs remains unclear. 

The boundary between the TN and Centaur regions and the dynamical definition of Centaurs are somewhat variable in the literature, and this hinders the comparison between different studies. Additionally, there are few observational estimates of the Centaur population from a well-characterized survey. On the contrary, TNOs and JFCs are much more observed and their size distribution and number are more constrained. Therefore, the predictions on Centaurs are mostly based on a TNO-JFC steady-state.   
The existence of a TN region as a source of JFCs was first suggested by \citet{Edgeworth1938} and \citet{Kuiper1951}; but it was not until the paper by \citet{Fernandez1980} that this matter was theoretically analyzed. In this work, Fern\'andez proposed a transneptunian belt between $\sim$35 and 50 au as an alternative, more efficient JFC source (compared to the Oort Cloud). Later, \citet{Duncan1988} addressed this problem through numerical simulations. The next two baseline papers that analyzed the TNOs as a source of JFCs were the studies by \citet{Levison1997b} and \citet{Duncan1997}. In the first paper the authors studied the evolution of 20 now called Classical Transneptunian Objects (CTNOs) plus clones, and found that some objects in their simulation crossed the orbit of Neptune and were scattered by this planet. Thus, they inferred the existence of an excited population in the TN region that could be an order of magnitude larger than the Classical Belt, which they called the Scattered Disk (SD). In the second paper, the authors suggested that the SD should produce more Ecliptic Comets (ECs) (a group that includes JFCs as well as Centaurs) than the Classical Belt because SDOs can approach Neptune during their perihelion passages and be scattered by this planet to orbits with shorter orbital periods.

As the structure of the TN region emerged thanks to the observations, four different sub-populations were identified. The classical transneptunian objects (CTNOs) with semimajor axes between  40 au $ \lesssim a \lesssim 50$ au and orbits with both low eccentricities and low inclinations, the resonant objects in mean-motion resonances (MMRs) with Neptune, such as the Plutinos in $2:3$ MMR, the scattered disk objects (SDOs) with perihelion distances between $30$ au $ < q \lesssim 39$ au that can cross the orbit of Neptune and eventually evolve into the planetary region becoming a Centaur, and the detached objects with $q > 39$ au that are decoupled from Neptune.

It became evident that the structure and dynamical characteristics of the TN sub-populations showed hints of a convulsed past. New planetary formation models based on planetary migration (e.g. \citet{Fernandezip84, Malhotra93, Malhotra95, Tsiganis05, Walsh11}; \\ \citet{Nesvorny12, Nesvorny15}) were necessary to explain the observations. Many studies have addressed the matter since the pioneering work of \cite{Fernandezip84}, where the authors found a radial displacement for Uranus and Neptune during their accretion and scattering of planetesimals, and the mechanism of radial migration came to light.  \cite{Malhotra93, Malhotra95} succeeded in explaining the capture of Pluto into the 3:2 MMR with Neptune, acquiring its high eccentricity and inclination from a migrating Neptune. Then, the Nice model \citep{Tsiganis05} opened the door to a series of works that focused on explaining the current orbital architecture of the Solar System. The planetary migration led, anyway, to a convulse early Solar System evolution in which a great mass depletion should have occurred. After this, the Solar System began to stabilize into the current form and dynamics, ultimately acquiring the present configuration. 

Once the stabilization had taken place, small body populations started interacting and established links between each other. In particular, Centaur objects are in a transient zone, permanently interacting with neighboring regions and populations. Their current population is mostly formed and defined by the contributions of their specific sources, mainly in the current TN region but also in other small body sources. A valuable review paper on 
formation, orbital properties, evolution and links between small body populations and reservoirs by \cite{Dones2015} would be helpful for an interested reader.

In this paper we are interested in analyzing the current Giant Planet Crossing population as a whole. This comprises objects with perihelion distances $q$ less than that of Neptune. However, Centaurs, in spite of being giant planetary crossers, are often defined more narrowly as objects with semimajor axis $a$ between those of Jupiter and Neptune. Thus, they do not represent the complete population of giant planetary crossers. In this work we define and address:
\begin{itemize}
    \item Giant Planetary Crossers (GPC): those objects with $5.2 $ au $< q < 30$ au.
    \item Centaurs: objects with $5.2 < a < 30$ au.
\end{itemize}

The limit of $q = 5.2$ au in the definition of GPC is due to the fact that in the region interior to the orbit of Jupiter, i.e. the Jupiter Family Comets (JFC) zone, the perturbations of the  terrestrial planets are necessary for the study of the dynamical evolution of small bodies. Besides, a physical model is also required to account for sublimation \citep{Disisto09}. All these factors prevent the study of Jupiter crossers to be handled in the same way than that of the other Giant planets.

In this paper we address the present day GPC, their dynamical evolution, dynamical lifetimes and their number and source feeding regions. In the next section we address the observed GPC and Centaur population, their observational features and size distribution and perform a dynamical evolution. In Sect. \ref{sd} we perform a numerical simulation of the evolution of SDOs and their contribution to GPC and Centaurs and in Sect. \ref{otrasfuentes} we address other secondary sources. In sect. \ref{todo} we join together all the contributions to the current GPC and Centaur population, and in the last section we present the conclusions.

\section{The observed GPC and Centaurs}
\label{obs}

\subsection{From surveys}
\label{survey}

The first Centaur to be discovered was (2060) Chiron by \cite{Kowal1977} on November 4, 1977; the first detection of a moving object between the giant planets. Thirteen years later, \cite{Meech1990} detected a low-surface-brightness coma for Chiron revealing for the first time the transitional nature of this object.
Chiron has an absolute magnitude $H= 5.8$ and an estimated diameter of $166$ km. It is the largest member of the population. The smallest Centaur discovered up to now is (2015 RK277) with $H = 15.5$. (5145) Pholus was the second Centaur to be discovered, in 1992, the same year that the first TNO was discovered \citep{Jewitt1992}. Many more were soon detected, but following surveys were focused more on finding TNOs than Centaurs. Therefore, in the present there are no surveys focused on finding only Centaurs. Instead, they generally are serendipitous discoveries from TNO designed surveys. As with all small body populations, we usually model the absolute magnitude ($H$) distribution of TNOs and Centaurs as following one or more exponential distributions including proposed breaks and or divots in that distribution. Considering the simplest form, as an exponential $N(H) \propto 10^{\alpha H}$, where $\alpha$ is the logarithmic slope which characterizes the population, it is possible to relate it with its size distribution. From the relation between magnitude and diameter: $D = 1327.5 \,\, 10^{ -H/5}/\sqrt{p_v}$, where $p_v$ is the albedo, the size distribution would result in a power differential law of the form: $N(D) \propto D^{-q}$, where $q = 5 \alpha +1$.
The first survey designed to discover and determine the orbits of hundreds of TNOs was the Deep Ecliptic Survey (DES) which worked from 1998 to 2005 observing in VR filter \citep{Elliot2005}. This  survey 
followed-up observations of 304 objects which allowed for well-determined orbits and dynamical classifications into the sub-populations of the TN zone, i.e. Classical, Scattered, and mean-motion resonances with Neptune and also Centaurs. \citet{Adams2014} accounted for the DES biases and estimated an exponential law magnitude distribution valid for TNOs and Centaurs which has a break at fainter objects. Based on observations of Centaurs which were made in the $7.5 < $ $H_r$  $< $ 11 range they inferred a broken law with $ \alpha_{1} = 1.02 \pm 0.01 $ for $H_r$  $ \lesssim 7.2 $ and $ \alpha_{2} = 0.42 \pm 0.02 $ for $H_r$ $ > 7.2 $. This gives a number of Centaurs with  $H_r$ $ \lesssim 7 $ of  $ 13 \pm 5 $.

A survey that has greatly increased the number of observed TNOs and
also Centaurs is the Outer Solar System Origins Survey (OSSOS), which 
operated between 2013 and 2017 and reported the discovery of 840 
objects \citep{Bannister18}. This survey observed in the r-band and in
a ``w'' wide-band filter, covering an area of 155 $\text{deg}^2$ of
sky to depths of $m_r = 24.1$–$25.2$, and was designed to discover TNOs with a careful quantification of the biases. This allowed the OSSOS team to develop a survey simulator to account for OSSOS biases 
\citep{Lawler2018a}. \cite{Lawler2018b} used a sample of 68 scattering
TNOs (defined by $a > 30$ au) and Centaurs ($a<30$ au), discovered 
mainly by OSSOS, to explore their $H$-distribution by using r-band 
observations. They assumed that scattering  TNOs and Centaurs are part
of a dynamically ``hot'' population with a common origin, since their 
orbits have been excited to higher inclinations and eccentricities by 
scattering off Neptune or past/current entanglement with mean-motion 
resonances. In this sense, their sample of Centaurs plus scattering 
TNOs has a common size distribution which could be different than 
those of the other TN populations, since they have a different 
formation and collisional evolution. This selection of objects allowed
also to be sensitive to a much fainter $H_r$ (i.e. smaller sizes). In 
fact, they have $H_r$ values of 6 to 14.5 due to the very close 
pericenter distances of some of the TNOs in the sample.
\cite{Lawler2018b} obtained that both a divot and a knee distribution (see e.g. Fig. 2 in \cite{Shankman2016}) fits the data. Their preferred knee distribution has the knee in  $ H = H_b = 7.7 $ with  $\alpha_{1} = 0.9$ for $H < H_b$ and $\alpha_{2} = 0.4 $ for $ H > H_b$, while their preferred divot distribution transitions from bright- ($\alpha_{1} = 0.9$) to faint-end ($\alpha_{2} = 0.5$) slopes at  $H_b = 8.3 $ with a divot contrast $c = 3.2$. By using this last distribution, the authors used the  OSSOS Survey Simulator to determine the number of SDOs and Centaurs brighter than a given magnitude H and estimating an SDO population of $N_{SDO} (H_r < 12) = 2.7 \pm 0.7 \times  10^{6}$  (which corresponds to $D \gtrsim 20$ km for an albedo of 0.06), and $N_{SDO}(H_r < 8.66)= 9 \pm 2 \times 10^4$ ($D \gtrsim 100$ km). Assuming  that the size distribution holds for smaller sizes, the authors found that $N_{SDO}(H_r < 18)  =  3 \times  10^9$ ($D \gtrsim 1$ km). 
For Centaurs: $N_{C}(H_r < 12)= 3500 ^{+1800} _{-1400}$, and $N_{C}(H_r < 8.66)= 110^{+60}_{-40}$.
However, it should be noted that the size distribution at smaller sizes is unknown. Observational analysis of the size distribution of small craters on the satellites of the outer planets have shown that the size distribution of the impactor population could have a new break (see \cite{Bierhaus2015} for a good summary). This possible break was found in particular in the satellites of Saturn  \citep{Kirchoff2010} and on Pluto and Charon \citep{Singer19} from Cassini and New Horizons observations.

\cite{Kirchoff2010} obtained differential slopes  from $\sim 2.1$ to $3.67$  for small craters on the mid-sized Saturnian satellites.
 \cite{Singer19} found that crater data from Pluto and Charon indicate a shallow differential slope with $q \sim  1.7 $ for craters from $\sim 1$ to 13 km in diameter corresponding to impactors from $\sim 100$ m to 1 km in diameter. Small craters observed on the TNO Arrokoth by the New Horizons flyby \citep{Spencer2020} are also consistent with the slopes seen for small craters in the Pluto system.

\cite{Nesvorny2019} used previous models of Solar System evolution with a slow, long-range and grainy migration of Neptune to predict the orbital element distributions of current Centaurs by testing the models using the OSSOS survey simulator. They obtained a good match to the observed OSSOS Centaur orbital distribution. 
The size distribution of the primordial outer disk was calibrated from Jupiter Trojans and, after running the OSSOS survey simulator, the authors also obtained a good match between their model and the observations. Therefore, they predicted a population of Centaurs of $ 21, 000 \pm 8, 000$ for $D > 10$ km with a size distribution that can be obtained from that of Jupiter Trojans ($N(>D) \propto D^{-2.1}$ for $5 < D < 100$ km). From this distribution, the number of Centaurs with $H_r < 12$ ($D \gtrsim 20$ km) would be  $\sim 4,900 \pm 1,800$,  which is in agreement with the estimation of \cite{Lawler2018b}.

The number and size distribution of Centaurs is strongly linked to the number and size distribution of SDOs (i.e. their source in the SD), which can be considered as a sign of their origin. 

\subsection{Dynamical evolution of observed GPC}
\label{evolution}

\begin{figure}[h!]
 \includegraphics [width=1.\textwidth]{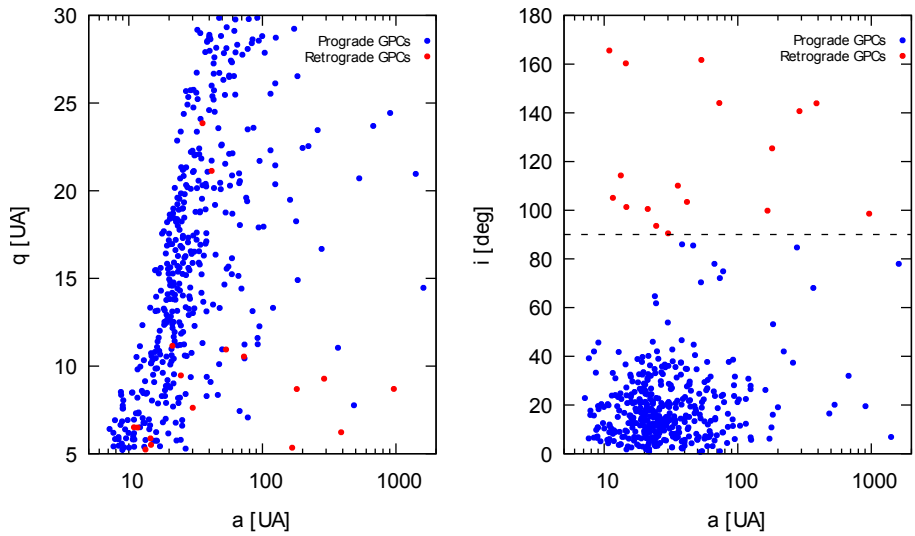}
\caption{Semimajor axis vs perihelion (left panel) and semimajor axis vs inclination (right panel) of the catalogued GPCs up to May 1st 2019.}
\label{obscentfig}
\end{figure}

The observed population of objects in the Giant Planet zone has greatly increased in recent years. 
In order to study and quantify the evolution of the current observed giant planetary crossing population, we developed a numerical study of its dynamical evolution. 
There are 432 GPCs ($5.2$ au $< q < 30$ au) listed in the Minor Planet Center database up to May 1st, 2019. Of these objects, 17 have retrograde orbits and 415 have prograde orbits. The orbital element distribution of all the objects is shown in Fig. \ref{obscentfig}.

For the simulation we consider these 432 GPCs and calculate the orbital parameters at the common epoch of 2019 April 27th for those few objects whose data differed from this epoch. 
For each catalogued GPC we build six clones, replacing its original mean anomaly by random values between 0$^{\circ}$ and 360$^{\circ}$ and  perform a numerical integration considering a total of 3024 particles (432 real plus 2592 synthetic) over the age of the Solar System. We use the hybrid integrator EVORB \citep{Fernandez2002} with an integration step of 0.2 yrs to follow the dynamical evolution of the particles under the gravitational influence of the Sun (including the masses of the terrestrial planets), the four giant planets and Pluto. Each particle evolves for 4.5 Gyr unless removed from the simulation due to a collision with one of the outer planets, reaching a semimajor axis $a >$ 5000 au or a position with $r <$ 5.2 au, i.e. entering the JFC zone where the perturbations of the terrestrial planets are not negligible and a physical model is also required to account for sublimation \citep{Disisto09}. In addition, we exclude the plutinos from the simulation.

From the initial 3024 particles, the results of the simulation show that 3 particles collide with a planet (1 with Uranus and 2 with Neptune). 1748 particles (57.8\%) reach a $>$ 5000 au where they are considered ejected from the Solar System, 1265 (41.8\%) become JFCs and 8 particles survive the total integration time. In addition, we computed the total number of individual encounters between the particles and the outer planets if the particles reached a distance to the planet within their Hill radii. We found that of the total number of encounters with the massive bodies, 0.15\% are with Pluto, 63.51\% are with Neptune, 27.68\% with Uranus, 8.61\% with Saturn and 0.05\% with Jupiter. We further calculated the percentages of the total number of particles that have encounters with the massive bodies during their evolution, obtaining that 22.19\% of the particles have one or more encounters with Pluto, 81.61\% with Neptune, 81.55\% with Uranus, 67.43\% with Saturn and 8.6\% with Jupiter.

\begin{figure}[h!]
 \includegraphics [width=1\textwidth]{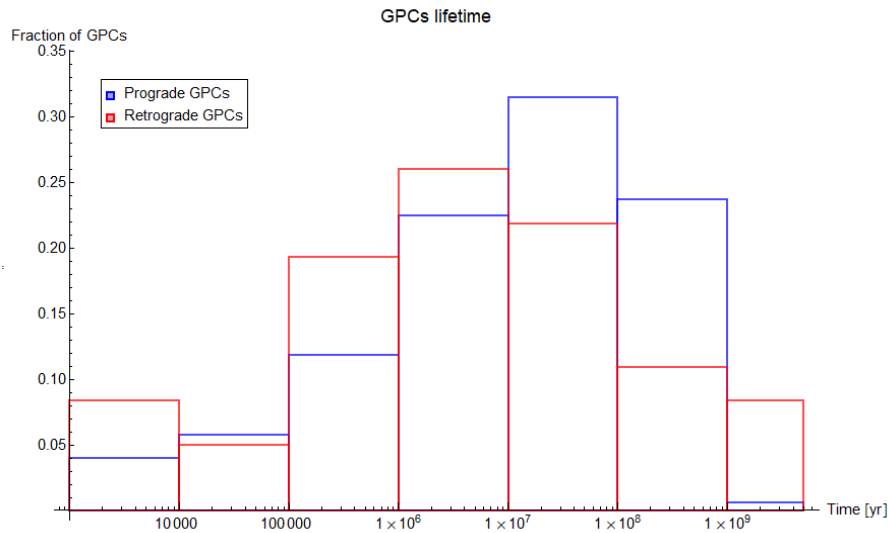}
 \caption{Normalized distributions of the GPCs lifetime for both prograde and retrograde objects.}
\label{histor}
\end{figure} 

Throughout the complete simulation time we recorded the orbital elements of all the described particles every $1,000$ years. Our results show that all the integrated GPCs have a median dynamical lifetime of $\sim$ 16.37 Myr in the GPC zone. 
In Fig. \ref{histor} both the retrograde and prograde GPCs dynamical lifetimes are shown as normalized distributions. The retrograde GPCs have a median dynamical lifetime of $\sim$5.48 Myr, which is much shorter than the median dynamical lifetime of $\sim$17.43 Myr of the prograde GPCs. This difference may indicate that the observed retrograde GPCs experience a faster evolution than the observed prograde GPCs. This could be connected with the different perihelion distribution of the observed prograde and retrograde GPCs in our sample. Indeed, it is seen from Fig. \ref{obscentfig} that most of the retrograde objects (15 of 17) have perihelion distances of less than $\sim  12$ au. \cite{DiSisto2007} noted  that the dynamical  evolution in the Giant planetary zone is strongly dependent on the perihelion distance, being the lowest perihelion objects those who evolve faster. Therefore, it is probable that the faster evolution of our initial retrograde GPCs is due to the fact that the majority of them have low perihelion values with respect to those of the prograde population. The dynamical evolution of the two retrograde objects with high perihelion distances, i.e. the cases of 2011 KT19 and 2008 KV42, have been studied by \cite{Chen2016} who obtained a mean dynamical lifetime of 500 Myr or greater and \cite{Gladman2009} and \cite{Brasser2012} who obtained a median lifetime of 200 Myr for 2008 KV42. However, \cite{Brasser2012} claimed that the Oort cloud dominates over the SD as a source of the population with $i >70^{\circ}$, $15$ au $ < q < 30$ au and $a < 100$ au, and found that those objects remain with their perihelion pinned to Uranus for a long time, showing short-period oscillations related to the Kozai mechanism. 
We found similar dynamical behaviors in our simulation for both long-lived objects.

Previous studies have been performed on the observed GPCs. Tiscareno and Malhotra (2003) explored the long term dynamical behavior of 53 known GPCs as of May 1st 2002 and followed their evolution for 100 Myr considering the perturbations of the four outer planets. They found that the median dynamical lifetime of their GPCs in their simulation was of 9 Myr and concluded that their sample was probably biased towards low eccentricities and low inclinations. Their sample of GPCs differed from ours, which covers a wider range of semimajor axis and inclination. In addition, our simulation lasts for 4.5 Gyr while that of \cite{Tiscareno2003} lasts for 100 Myr, at the end of which 20\% of their sample have lifetimes exceeding 100 Myr. The combination of these changes in the simulation conditions could account for the difference we obtain in the median lifetime value.

\begin{figure}[h!]
 \includegraphics [width=1\textwidth]{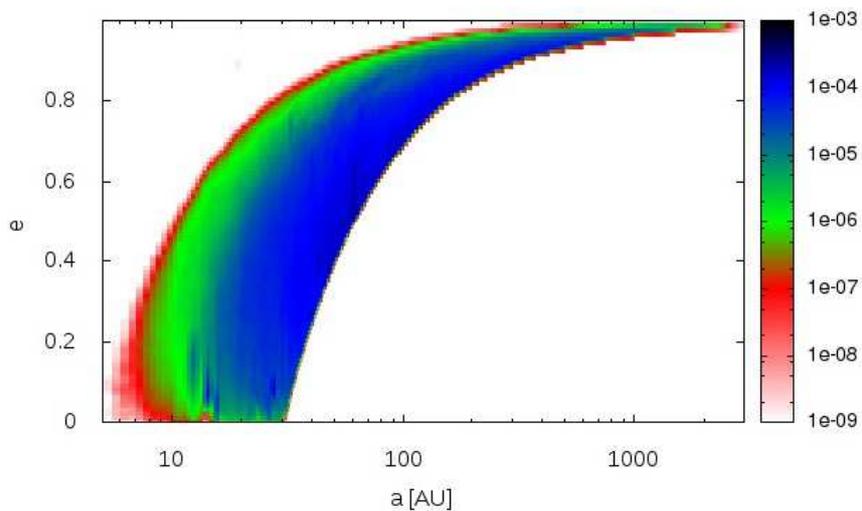}
 \caption{Time-weighted distribution of the integrated GPC in the semimajor axis (a) vs. eccentricity (e) space.}
\label{mae}
\end{figure}

\begin{figure}[h!]
 \includegraphics [width=1\textwidth]{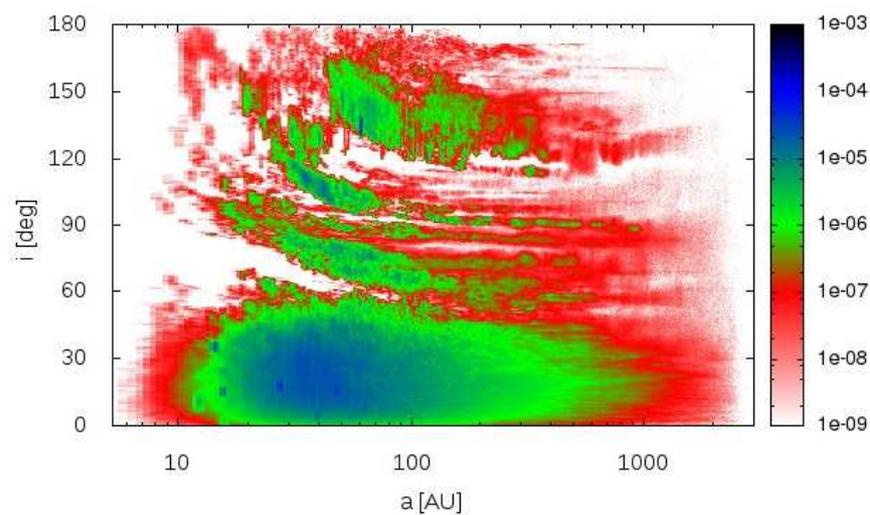}
\caption{ Time-weighted distribution of the integrated GPC in the semimajor axis (a) vs. inclination (i) space.}
\label{mai}
\end{figure}

\citet{Tiscareno2003} found that two thirds of their simulated particles were ejected from the Solar System during their simulation while one third was injected into the JFC population, in agreement with our simulation. They also found that GPCs did not stay in resonances for more than a few Myrs, in contrast with the SDOs behavior. In our simulation we find that some particles were captured in MMR for long periods of time. These type of captures, in MMR at high eccentricities and large semimajor axis values, are also found in SDOs. As mentioned, there are eight particles that survive for all the 4.5 Gyr in our simulation, six progrades and two retrogrades. The prograde particles that survived experienced captures in MMR for long periods of time that could be of the order of Gyrs and also resonance sticking. The two surviving retrogrades show a combination of temporary long-term resonance locking or resonance sticking moving from one MMR resonance to another and also Kozai resonances, and a conservation of perihelion distance in the zone between Saturn and Neptune but avoiding close encounters with the planets. The resonance sticking mechanism was first noticed by \cite{Duncan1997} and since then is has been studied by a number of authors. In particular, it was found for objects in Neptune encountering orbits (scattered TNOs) \citep{Lykawka2007, Yu2018}, and in the temporarily capture of Uranian and Neptune Trojans from the dynamical evolution of Centaurs \citep{Alexandersen2013}. We note several resonance captures in the whole evolution  with a tendency of GPCs to be captured into MMR with Neptune: 1:1, 3:2, 2:1 and 3:1 and 5:2. This can be noted in Figs. \ref{mae} and \ref{mai} where we plot the normalized time-weighted distribution for the GPC in the orbital element space. These maps show the normalized time fraction spent by the observed GPC across their evolution in different regions of the (a, e, i) space.

The blue zones are the most visited and thus, the ones were GPCs spend most of their time. Therefore, there is a dependence of the time of permanence on perihelion distances. The evolution near Neptune is slower, meaning slower diffusion timescales, and becomes faster close to Jupiter (see Fig. \ref{mae}). 
In Fig. \ref{mai} the different dynamical evolution of prograde and retrograde GPCs can be seen. The densest zone (that of higher permanence) can be seen clearly for inclinations below $45^{\circ}$, while high inclination objects experience a faster evolution. However, there are some blue features visible in the retrograde population zone showing MMR captures as well.


\section{The GPC and Centaur populations from the Scattered Disk}
\label{sd}

The giant planet region is a transitional zone which is mainly crossed by incoming TNOs, in particular those from the SD. In this paper we present new calculations about the current contribution of SDOs to the zone that comprises orbits with perihelion distances less than that of Neptune, i.e. the current population of giant planet crossers. \citet{DiSisto2007} (DB07 in the following) built a model of an intrinsic current SD with real SDOs plus clones, which were numerically integrated for $4.5$ Gyr to follow their dynamical evolution. They  calculated the SDO contribution to the GPC zone ($q < 30$ au) in contrast to SDOs that have $q > 30$. 
\citet{DiSisto2007} obtained a rate of injection of SDOs to GPC of $5.2 \times 10^{-10} yr^{-1}$ and a number of GPC with radius $R > 1$ km equal to $2.8 \times 10^{8}$. \cite{Volk2008} also built a debiased model of the orbital distribution of the SD, and numerically integrated the particles for 4 Gyr. They defined the  SDOs as those TNOs with $ q > 33$ au and $a > 50 $ au and found an escape rate of $1 - 2 \times  10^{-10} SDOs/yr$, lower than DB07, mainly due to the difference between both debiased semimajor axis distributions. They also found that $10^6$ Centaurs with $D > 1$ km must exist in order to balance the loss of JFCs. 

Since the SD is the main source of Centaurs and in light of the new observations and constraints on the size distribution of TNOs, we perform new numerical simulations following the model by DB07 but  updated with the current observed SDO population. 

\subsection{The model}
\label{model}

Since the model developed by DB07 in 2007, the number of observed SDOs has greatly increased. \citet{DiSisto2007} recorded 95 observed SDOs from the Minor Planet Center database while  now (up to April, 2019) this number has risen to 603. Those SDOs were defined as TNOs, which are not plutinos, with perihelion distances $ 30 < q < 39$ au, semimajor axis $a > 40$ au and eccentricities $e>0.2$ to distinguish them from CTNOs \citep{Elliot2005}. Those SDOs have absolute magnitudes in the range: $ -1.1 < H < 9.8$, being (136199) Eris ($H = -1.1$) the largest member with $D \sim 2500$ km and 2015 PK312 ($H = 9.8$) the smallest observed object with $D \sim 50$ km.

DB07 developed a model of the current SDO population which accounts for the bias in the discovery probability for different semimajor axis (based on \cite{Fernandez2004}) since SDOs can be discovered when they are close to their perihelia. Therefore, they found that  
an intrinsic semimajor axis distribution of SDOs would be given by: 
\begin{equation} \label{fdea}
f(a)  \propto a^{-2}.
\end{equation}

DB07 also considered the bias towards low inclination discoveries by existing surveys, and proposed  a Brown intrinsic inclination distribution \citep{Brown2001} given by:
\begin{equation} \label{fdei}
F(i) di \propto \sin i \exp ^{\frac{-i^2}{2 \sigma_{i}^{2}}},
\end{equation}

 where $\sigma_i = 12^{\circ}$ from the model developed by \cite{Morbidelli2003}. From observational surveys, recent works have found different values of $\sigma_i$. \cite{Gulbis2010} analyzed the inclination distributions of the different TN populations based on data from the Deep Ecliptic Survey. They found that for the scattered population, the general inclination distribution form is  consistent with a Brown distribution with $\sigma_i = 6.9^{\circ}$. \cite{Petit2017} study the High Ecliptic Latitude extension (HiLat) of the Canada–France Ecliptic Plane Survey (CFEPS) and found that for other populations than the classical one, the combined CFEPS + HiLat sample allows an inclination distribution with $12.5^{\circ} < \sigma_i < 20^{\circ}$. Therefore, although there is disparity in the survey $ \sigma_i$ values, the adopted value of  $\sigma_i = 12^{\circ}$ seems to be consistent with the observations.

\begin{figure}[h!]
 \includegraphics [width=1\textwidth]{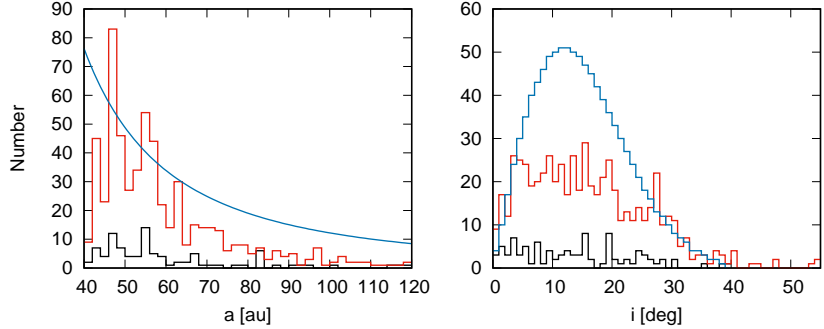}
\caption{Intrinsic orbital element distribution by DB07 (blue line), Observations used in DB07 model (black line), observations up to April, 2019 (red line).}
\label{testfig}
\end{figure}

Our first test for the present review was to study if considering the new observations, the orbital distributions of observed objects approaches the intrinsic distribution proposed in DB07 model. This comparison can be seen in Fig. \ref{testfig}.
As shown in the figures, the current observations seem to follow the same trend as the debiased semimajor axis and inclination distributions. In the updated observations, there are some objects with higher inclinations and others at larger distances than  the observations used in DB07, but they account for a small fraction of the total number of objects. This analysis encouraged us to perform new numerical simulations with the model used in DB07 but considering all the updated observations. Therefore, in the present work we use Eqs. (\ref{fdea}) and (\ref{fdei}) to build the model as we explain in the next section.

\subsection{The numerical simulation}
\label{simulation}

\begin{figure}[h!]
 \includegraphics [width=1\textwidth]{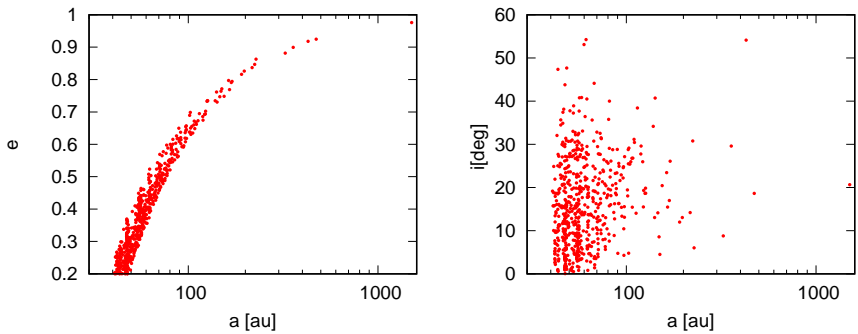}
\caption{Semimajor axis vs eccentricity and Semimajor axis vs inclination of the real SDOs by April 2019.}
\label{obsfig}
\end{figure}

As mentioned, our current observed population of SDOs ($ 30 < q < 39$ au, $a>40$ au and $e>0.2$) is of 603 objects, up to  April, 2019. They are plotted in Fig. \ref{obsfig}. 

With those real SDOs (603) we generate 5167 clones so that the total of the initial particles (5770), that is to say real plus clones, fit the distribution of semimajor axis (given by Eq. (\ref{fdea})) and inclination (given by Eq. (\ref{fdei})) of the model. We proceed in the same way that in DB07. We first generate the number of clones for each value of semimajor axis, then change the semimajor axis of almost half of the clones (selected randomly from the total sample of synthetic SDOs) by $\delta$  such that $-2\times10^{-4} < \delta < 2\times10^{-4}$ and allocate the mean anomaly for all the clones randomly between $0^\circ$ and $360^\circ$. Then, we assign random inclinations for the clones in such a way that the total number of particles (real plus clones) fit the proposed Brown distribution (Eq.\ref{fdei}).

Therefore, we performed  a numerical integration  of 5770 massless particles under the gravitational influence of the Sun (including the masses of the terrestrial planets), the four giant planets and Pluto, with the hybrid integrator EVORB \citep{Fernandez2002}. Pluto was only included in the simulation for the purpose of a future work. Its
effect on the dynamical evolution of SDOs and their link to GPC and Centaurs is negligible as was previously noted by \cite{Nesvorny2000, Tiscareno2009}; even the effects of the largest TNOs are minor in the supplying of Centaurs and JFCs \citep{Munoz2019}.

The  integration step was $0.2$ years and the total time span, $4.5$ Gyr. The code cutting conditions were: collision with a planet, reaching a semimajor axis $a > 5,000$ au (ejection), or a distance $r<5.2$ AU, i.e. the zone of Jupiter Family Comets (JFC) where  the terrestrial planets perturbations are needed to account for a real dynamical evolution and a physical model is also required \citep{Disisto09}.

\subsection{Results}
\label{results}

The general results of the new simulation are similar to that of DB07. However, in the new simulation we include a larger number of real objects which extent to larger $a$ and $i$ values, the integration step is somewhat smaller than in the previous integration and we include Pluto as another perturbing massive object. There are also new estimations of the number and size distribution of SDOs. Therefore, we update the results regarding the contribution of SDOs to the GPC population and also evaluate their contribution to Centaurs. 

From the initial 5770 particles, 18 ($0.3 \%$) collide with a planet: 4 with Saturn, 4 with Uranus and 10 with Neptune. 3801 particles ($50.6 \%$) reach $a > 5000$ au, 884 ($15.3 \%$)  reach the zone of $r  < 5.2$ AU, and the remaining 1951 ($33.8 \%$) survive as a SDO. These percentages are a little different from DB07; in particular the number of objects that enter the JFC zone is lower while the number of surviving particles is higher. This is due to the fact that in the present model we extend the SD up to much higher distances. Thus, the evolution of those particles is slower, having less chances of encountering Neptune,  and then tend to remain longer in the SD. On the other hand, as was found by DB07, the delivery to the JFC region is mainly from regions with small perihelion distances and small semimajor axes. 

We computed the encounters between a particle and a planet if the particle reached a distance to the planet within their Hill radii.
From the total number of encounters,  $4.9\%$ are with Pluto,  $73.4 \%$  with Neptune, $17.8 \%$ with Uranus, $3.8 \%$  with Saturn and $0.1 \%$ with Jupiter. From the total number of particles in the integration, almost all, i.e. $92 \%$ encounter Pluto,  $46 \%$ encounter Neptune, $38 \%$ encounter Uranus, $28 \%$ Saturn and $4 \%$ Jupiter. However, the encounters with  Jupiter and, in particular, the number of particles at encounter, is sub-assessed due to the cutting condition at $r = 5.2$ au.
It has to be noted that, although the majority of the particles encounter Pluto, the real proportion of encounters with Pluto in the whole simulation (12 per particle) is much smaller than that with Neptune (358 per particle) which is the planet that actually scatters the SDOs towards the GPC and Centaurs region. The encounter statistics give a mean probability of encounters inside the Hill radius per object per year of $2.7 \times 10^{-9}$ for Pluto and $8 \times 10^{-8}$ for Neptune, more than an order of magnitude larger than for Pluto. Those numbers are in agreement with the probability calculated by the Opik method. Therefore, our simulation confirms that Pluto has a minimal effect on the supply of GPC, Centaurs and therefore JFCs, as was previously stated \citep{Nesvorny2000,Tiscareno2009}.

\begin{figure}[h!]
 \includegraphics [width=0.95\textwidth]{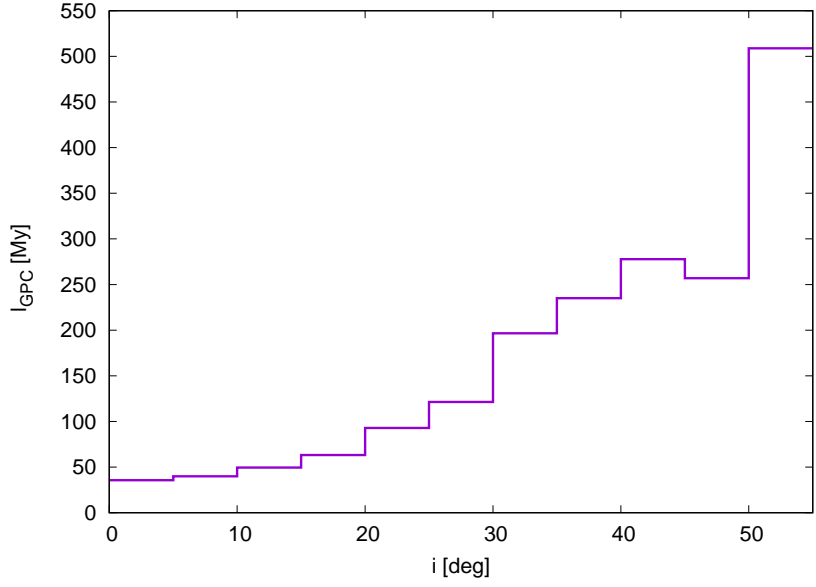}
\caption{Mean lifetime as a GPC vs initial inclination.}
\label{ltiifig}
\end{figure}

From the initial SDOs in the integration, $70 \%$ are delivered to the GPC zone. 
The mean lifetime ($l_{\text{GPC}}$) there is $68$ Myr, the most likely value being between 10 and 100 Myr as was found in DB07 (see Fig. 4). There is a strong dependence of the lifetime as a GPC with the initial inclinations in the SD as can be seen in Fig. \ref{ltiifig}. It is worth noting that SDOs with initial inclinations between 50 and 55 degrees have large mean lifetimes in the GPC zone. However from the Brown distribution they are a few number of particles. 
Also, we notice the same heavy reliance of mean lifetime of GPC with the perihelion distance as DB07 (see their fig. 6), the mean lifetime being larger for greater q values. 

\begin{figure}[h!]
 \includegraphics [width=0.95\textwidth]{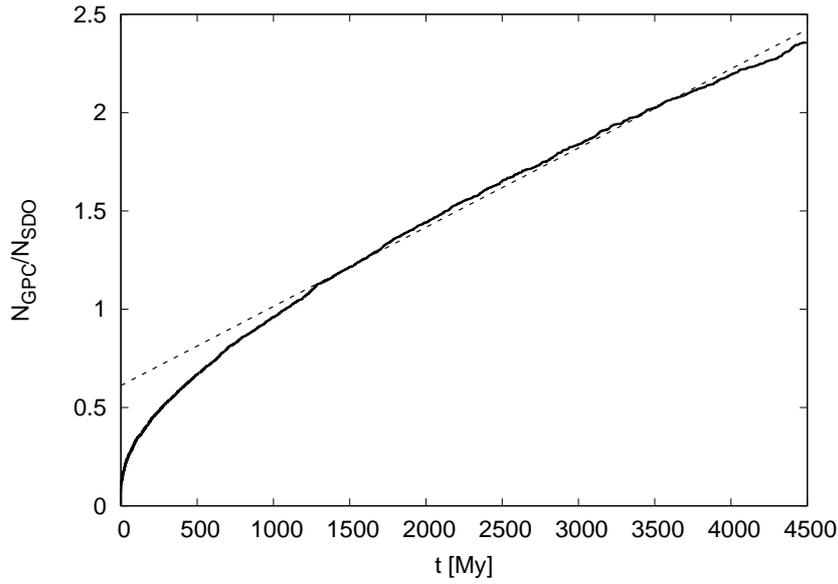}
\caption{Number of GPC ($N_{GPC}$) coming from the SD with respect to the surviving SDOs ($N_{SDO}$) as a function of time (solid line). The dashed line represents the linear fit to data (see text).}
\label{tasafig}
\end{figure}

From the simulation we estimate the injection rate of GPC from the SD. We recorded every 1,000 years, the orbital elements of SDOs when they have $q< 30$ AU. From this file, we compute the first time a SDO enters this zone as the time of injection. Therefore, we can analyze the time dependence of the quotient between the cumulative number of SDOs that enter the GPC zone ($N_{GPC}$) and the number of SDOs that remain in the SD ($N_{SDO}$). This relation is plotted in Fig. \ref{tasafig} and is well fitted by the linear relation whose slope is:
\begin{equation} 
\label{tasainy}
d[N_{GPC} / N_{SDO}]/dt = Y
\end{equation}

where  $Y  = 4.025 \pm 0.008 \times  10  ^{-10}  N_{SDO}  /year$  is  the rate of injection of GPC from the SD.
This linear fitting implies that if we know the current number of SDOs, the present cumulative number of  GPCs ($N_{GPC}(>D)$) coming from the SD would be given by:
\begin{equation} \label{npc}
N_{GPC}(>D) = Y \,\, N_{SDO}(>D) \,\, l_{GPC},
\end{equation}
where $N_{SDO}(>D)$  is  the current population of the SD greater than a diameter $D$, and the rate of injection will be $ Y \,\, N_{SDO}(>D) /year$. 
There are various estimations of the SD population from different surveys. \cite{DiSisto2011} analyzed the number and size-frequency distribution (SFD) of SDOs based on the works of \cite{Parker2010a, Parker2010b}. They adopted a broken power-law size distribution 
with a differential index of large objects given by  $s_1 = 4.7$ \citep{Elliot2005}, a break at diameters $d \sim 60$ km and two limit values for the differential index $s_2 = 2.5$ and $3.5$ for $d < 60$ km 
given the uncertainty of the SFD for small objects \citep{Bernstein2004, GilHutton2009, Fraser2009, Fuentes2008, Fuentes2009}. However, the recent discoveries by OSSOS increased the number of small SDOs and Centaurs and allowed for new estimations of the SFD and number of objects. \cite{Lawler2018b} found that a break in the SFD is required at $D \sim 100$ km. They found a faint-end slope of the magnitude size distribution $\alpha = 0.4 - 0.5$ which corresponds to a differential size index of $s = 3 - 3.5$.   
Therefore, we propose here the same SFD of SDOs as in \cite{DiSisto2011} but with the break at $D = 100$ km. Also, we consider the differential index for $D < 100$ km for three values $s_2 = 2.5, 3$ and $3.5$.
Thus, the cumulative number of SDOs will be given by:
\begin{xalignat}{4}
	N(>D) &= C_0 \bigg(\frac{1 \text{km}}{D}\bigg)^{s_2 - 1} &&\text{for} && D \leq 100~\text{km},  \nonumber \\
	N(>D) &= \text{3.5} \times 10^{5} \bigg(\frac{100 \text{km}}{D}\bigg)^{s_1-1} &&\text{for} && D > 100~\text{km},
	\label{nr}
\end{xalignat} 
where $C_0 = \text{3.5} \times 10^{5} \times 100^{s_2-1}$ by continuity for $D$ $=$ 100 km. 

This relation is plotted in Fig. \ref{ncent}.  We obtained for the intermediate index $s_2=3$ that $N(>2$km)$ = 8.75 \times 10^{8}$ and  $N(>10 $ km$) = 3.5 \times 10^{7}$, while \cite{Nesvorny2017} obtained $N(>2 $ km$) = 4.4 \times 10^{8}$ and  $N(>10 $ km$) = 1.5 \times 10^{7}$ for the inner SD ($50 < a < 200$ au, i.e. a part of the complete SD) by calibrating the source population of the observed ecliptic comets through a dynamical model. Both estimations are in agreement with ours within a factor of $\sim 2$, which given the slightly different population ranges is considered acceptable.

It has to be noted that it has been inferred from cratering counts on the Pluto and Charon system \citep{Singer19} and from cratering studies on the satellites of the outer planets \citep{Kirchoff2010,DiSisto2013,Bierhaus2015} that there may be an additional break in the SFD slopes of TNOs below 1-2 km in diameter. A series of geological studies in relation to impact cratering has been performed particularly on the Pluto system  \citep{Moore2016,Robbins2018}. A discussion about how geologic processes affect the crater size distribution was carried out by  \citet{Singer19}. However, a general study about geological processes that could erode small craters on all the objects which are the targets of collisions by Centaurs would be recommendable before asserting the existence of a new break. Therefore, we are not going to consider this range of sizes in the following. 

The cumulative number of GPC ($N_{GPC} (> D)$)  can be calculated from Eqs. (\ref{npc}) and (\ref{nr}).
For example, we have 9560 GPC with $D > 100$ km and using the intermediate index $s_2 = 3$ for the SFD of SDOs for small objects $N_{GPC}( > 10 $km$) \sim 2.3 \times 10^6 $ and $N_{GPC} (> 1 $km$) \sim 2.23 \times 10^8$, a little less than what was found by DB07. The rate of injection, calculated by $ Y \,\, N_{SDO}(>D) /year$ yields for example, 3 SDOs with $D > 1$ km every 2 years or 14 SDOs with $D > 10$ km every 100 years.

\begin{figure}[h!]
 \includegraphics [width=1\textwidth]{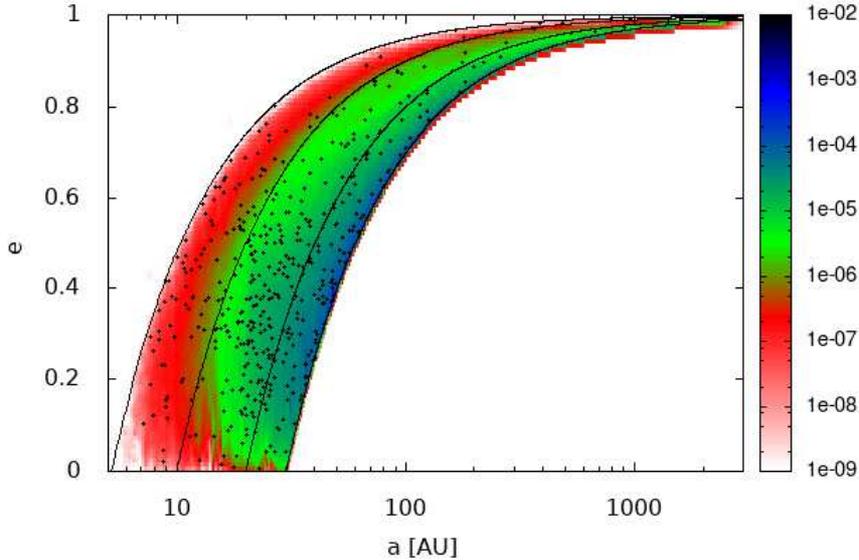}
\caption{Time-weighted distribution of the GPC obtained from the simulation in the semimajor axis (a) vs. eccentricity (e) space.  Black points represent the observed population. }
\label{mapaae}
\end{figure}

\begin{figure}[h!]
 \includegraphics [width=1\textwidth]{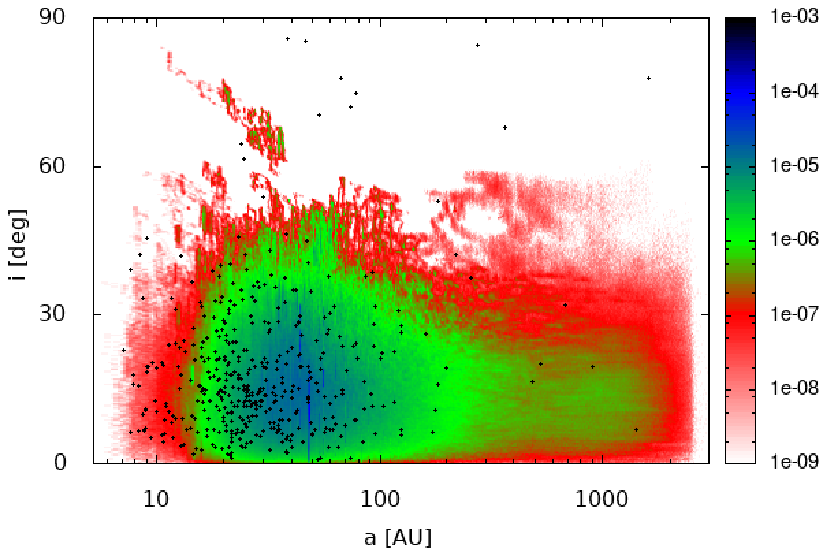}
\caption{ Time-weighted distribution  of the GPC obtained in the simulation in the  semimajor axis (a) vs. inclination (i) space.  Black points represent the observed population. }
\label{mapaai}
\end{figure}

From our numerical integration it is possible to calculate the normalized time-weighted  distribution for the GPC in the orbital element space. We divide the (a, e, i) space in bins of size $\delta a = 0.2$ au, $\delta e = 0.01$  and  $\delta i = 0.2^{\circ} $ and calculate the normalized time fraction spent by GPCs in different regions of the (a, e, i) space. These calculations are plotted in Figs. \ref{mapaae} and \ref{mapaai}. 
Those plots form dynamical maps of permanence and thus represent our modeled GPC distribution, assuming time-invariability. The observed GPC  are also plotted. The color code is indicative of the fraction of time spent in each zone (blue for most visited regions, red for least visited).
As in DB07, the orbital element region  with greater probability of occupation is the one with $\sim 20 \lesssim a \lesssim 80$ au, and $ i \lesssim  30 ^{\circ}$. In Fig. \ref{mapaai} 
there are various blue vertical features that correspond to mean motion resonances, being the $ 2:1$ at $a \sim 48$ au, corresponding to the Twotinos, the most populated one. Some of those features correspond to objects that are initially in a MMR, however they are a very small fraction of the whole set. We detected captures on several MMR with Neptune, Uranus and Saturn being the most frequent the 1:1, 2:1, 5:2, 3:1 MMR with Neptune and the 2:3 MMR with Uranus.

The observed GPC are in the zone also comprised by our model. However, there are high inclination objects and retrograde ones (not plotted in Fig. \ref{mapaai}) that cannot be explained by our model. Therefore, the source of retrograde GPCs is not the SD but the Oort cloud as proposed by other studies (e.g. \cite{Volk2013, Brasser2012, Nesvorny2017, Nesvorny2019}). 
From Fig. \ref{mapaae}, the zone of perihelion distances between 20  and 30 au  must  be densely populated, according to  our model. In addition, we note that the diffusion time scale is short for objects that reach perihelion distances within that of Saturn, in comparison to those that lie farther.  

However, the instability of the region near the orbit of Jupiter 
is a by-product of the boundary conditions of our model, which is not valid inside the region delimited by this planet.  

\begin{figure}[h!]
\includegraphics [width=0.95\textwidth]{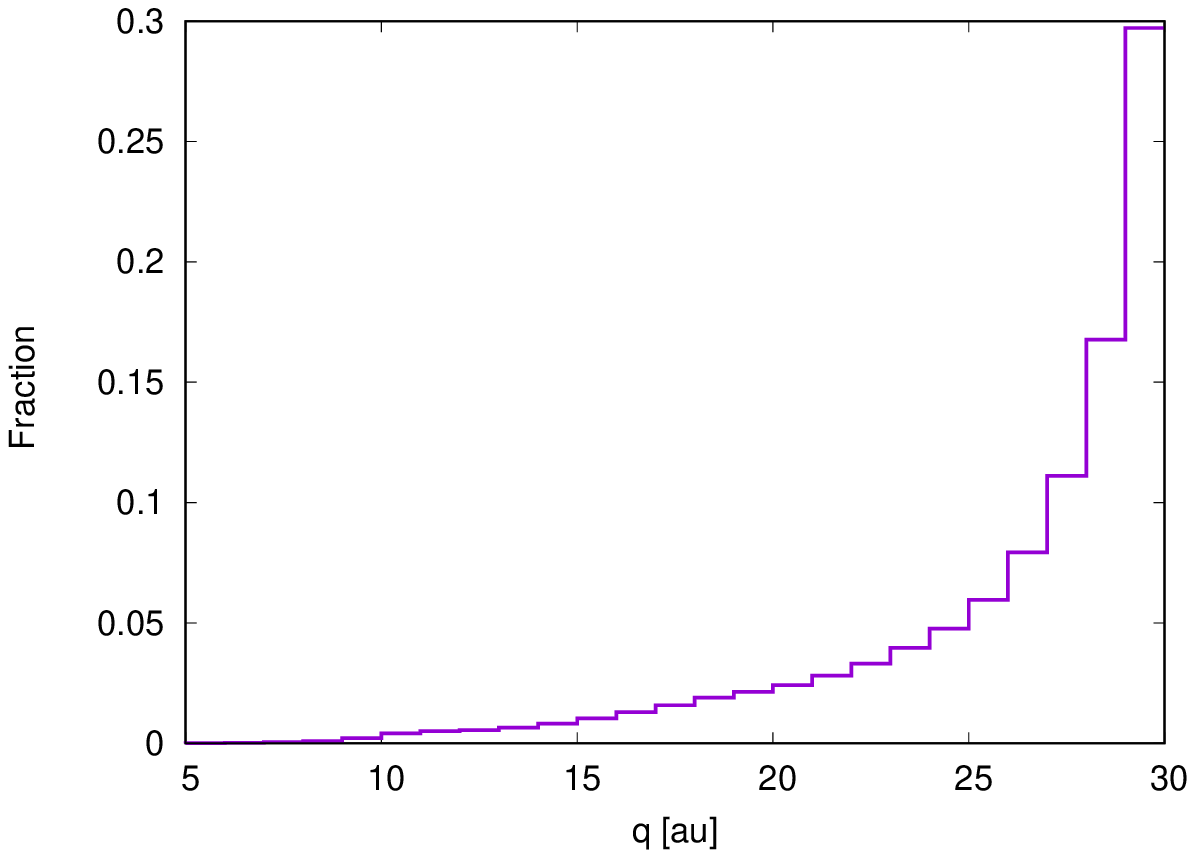}
\caption{ Normalized time-weighted distribution of the GPC obtained in the simulation with respect to perihelion distances $q$.}
\label{hqgpc}
\end{figure}

The dynamical evolution in the Giant planetary zone is strongly dependent on the perihelion distance, as is the mean lifetime in that zone. Therefore, the GPC are not uniformly distributed with respect to $q$. This can be seen in Fig. \ref{hqgpc} where we plot the normalized time-weighted  distribution of perihelion distances for the GPCs of our model. 
Rescaling this distribution  by $N_{GPC}(>D)$, obtained from Eqs. (\ref{npc}) and (\ref{nr}) it is possible to obtain the number of GPCs for each perihelion value. We note that the number of GPC grows exponentially with perihelion, being the region near Neptune the most populated one. This is connected with their origin as particles scattered by this planet. 

\subsection{The contribution to Centaurs}
\label{cent}
In the region restricted only by semimajor axes between $5.2 < a < 30$ au, the minor bodies are the Centaurs. There are observational results on the number and size distribution of Centaurs which can be used to compare them with our model. Although Centaurs are in general restricted to  $5 < a < 30$ au, our restriction on $a > 5.2$ au doesn't affect the time-averaged statistics, since the dynamical timescales get very short near Jupiter, as mentioned.

\begin{figure}[h!]
 \includegraphics [width=0.95\textwidth]{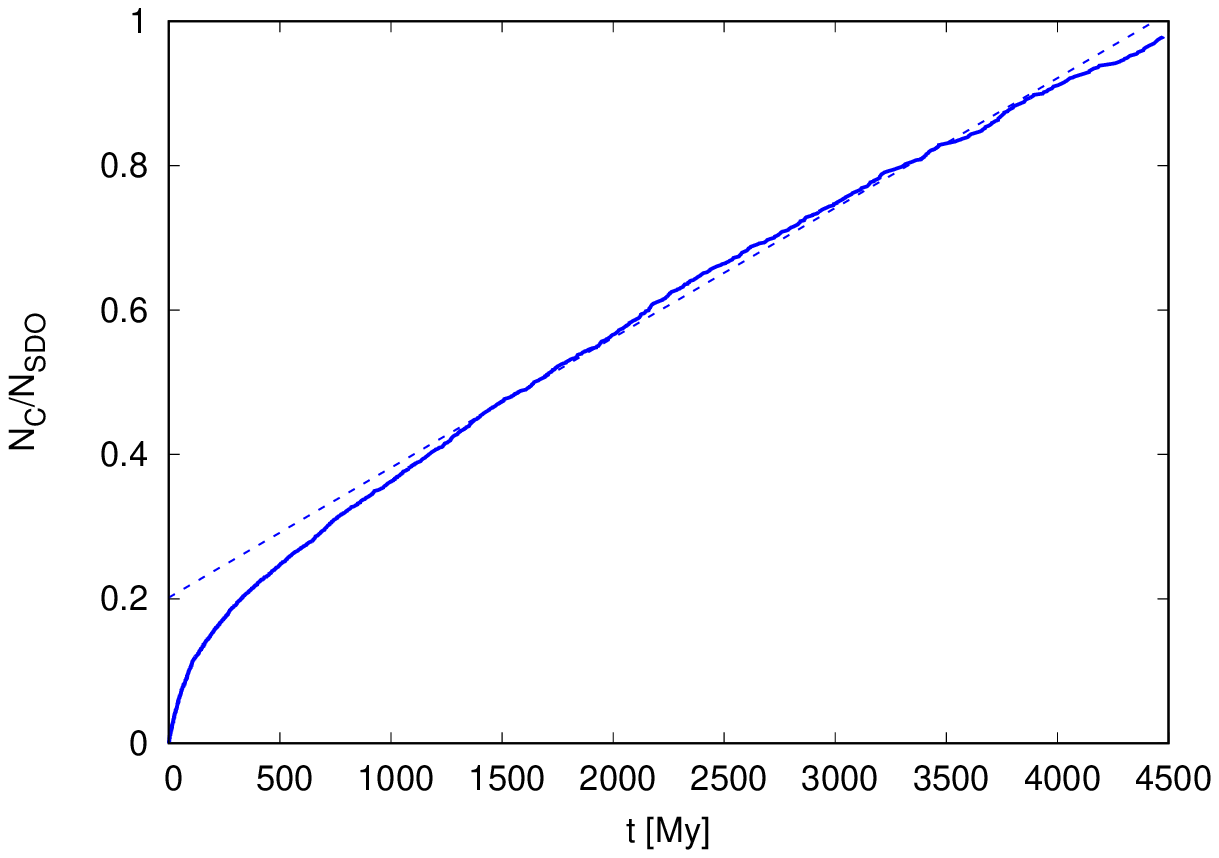}
\caption{Cumulative number of Centaurs ($N_{C}$) coming from the SD with respect to the surviving SDOs ($N_{SDO}$) as a function of time 
(solid line). The dashed line represents the fit to data (see text).}
\label{tasafigc}
\end{figure}

\begin{figure}[h!]
\includegraphics [width=0.95\textwidth]{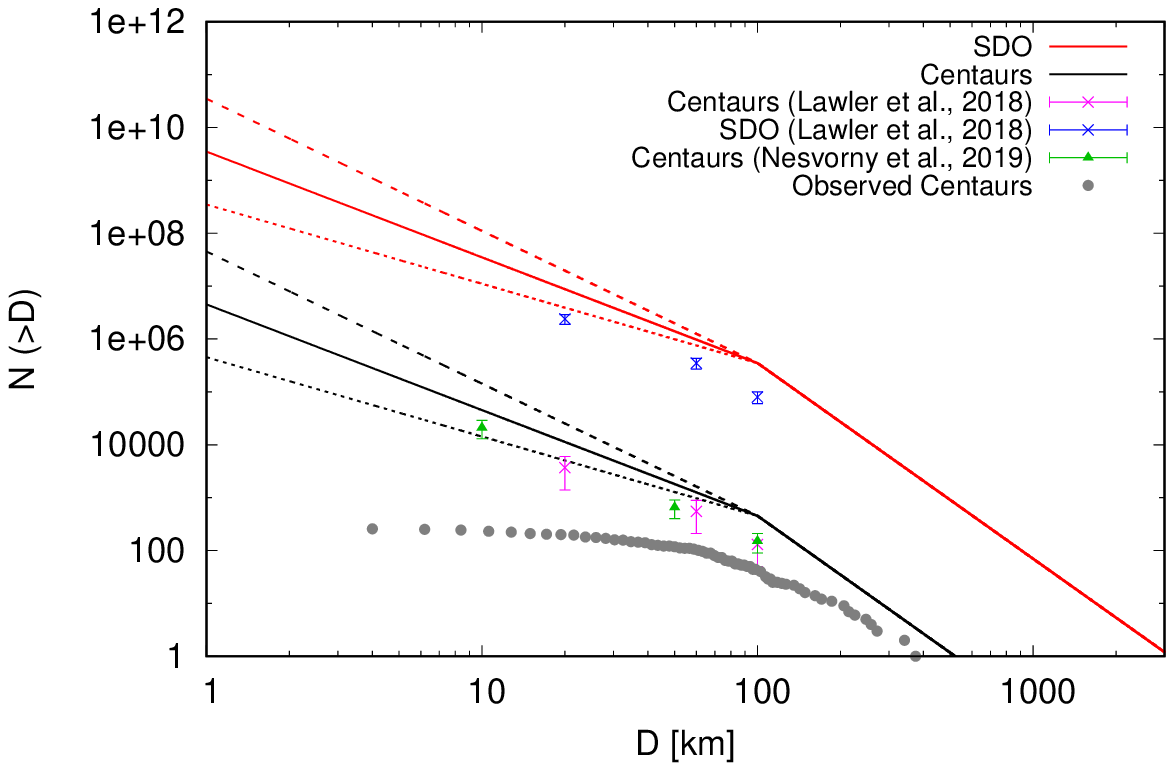}
\caption{Cumulative number of SDOs ($N_{SDO} (> D)$) obtained from Eq. (\ref{nr}) and cumulative number of Centaurs ($N_{C} (> D)$) from our model. Gray points correspond to the observed Centaur population, green points represent the number of Centaurs estimated by \cite{Nesvorny2019} and magenta points correspond to estimations of \cite{Lawler2018b}. Blue points represent the number of SDOs estimated by \cite{Lawler2018b}. For the observed population and survey estimations we use an albedo $p_v =0.06$ to convert absolute magnitude to diameter.}
\label{ncent}
\end{figure}

Thus, we  proceed as in the previous subsection, but considering the contribution of SDOs to Centaurs. $34 \%$ of SDOs enter the zone of $a < 30$ au, and we extract from the output orbit file, the first time a SDO enters this zone. We calculate the cumulative number of Centaurs  $N_{C}$ coming from the SD in relation to the surviving number of SDOs ($N_{SDO}$) for each time. 
 In this case, some SDOs left the integration, never entering the Centaur zone, and then we have to extract them from the remaining population at each time step.  This relation is plotted in Fig.  \ref{tasafigc} where we can see the linear fitting to the data. The slope of this line is the rate  of injection of Centaurs from the SD,
 $Y_C  = 1.796 \pm 0.005 \times  10  ^{-10} N_{SDO} /year$. 
The mean lifetime in the Centaur zone, calculated as before, is $l_C = 7.2$ Myr. Therefore, the cumulative number of Centaurs ($N_{C} (> D)$) from our model can be obtained from Eqs. (\ref{npc}) and (\ref{nr}) but considering $Y_C$, and  $l_{C}$ (instead of $Y_{GPC}$, and  $l_{GPC}$). 
From those calculations we plot in Fig. \ref{ncent}, the cumulative number of SDOs ($N_{SDO} (> D)$), the cumulative number of Centaurs ($N_{C} (> D)$) and two estimations  by \cite{Lawler2018b} and \cite{Nesvorny2019}. 
Our model predicts a somewhat greater number of large Centaurs than \cite{Lawler2018b} and \cite{Nesvorny2019}, but the numbers for $D < 100$ km are similar and lie between the error bars. \cite{Nesvorny2019} calibrated the Centaur population from Jupiter Trojans and predicted a population of Centaurs with a size distribution with a differential size index equal to $3.1$ for $5 < D < 100$ km, i.e. between our results for $s_2 = 3$ and  $s_2 = 3.5$. 

\subsection{Dynamical evolution}

The dynamical evolution in the GPC zone was analyzed from our new numerical simulation. Our results show that the particles that achieve the largest lifetimes exhibit essentially four types of dynamical behaviors, as presented in DB07. These are: 
\begin{itemize}
    \item Between the orbits of Saturn and Neptune, particles that present slight variations of $q$ throughout their whole lifetime,
    together with a nearly constant perihelion longitude, eccentricities exceeding $\sim 0.8$ and large semimajor axes ($>$ 100 au). This behavior tends to stabilize the orbit and its orientation, minimizing the number of encounters.
\item Transfers between mean motion resonances, known as resonance sticking, in the zone of $\sim 30$ au $ < a \lesssim 200$ au.  There are also transfers between MMR and Kozai resonances. 
\item For large values of the semimajor axis ($a \gtrsim 200$ au) the general dynamical evolution is a low increase of semimajor axis  but keeping the perihelion distances near Neptune. Thus, the objects are continuously entering and leaving the GPC zone up to ejection (injection to the GPC zone is more rare).   
\item GPC that enter a MMR or Kozai resonance and remain there up to the end of the integration. 

\end{itemize}

Those mechanisms are responsible of the long dynamical lifetime between $5.2 < q < 30$ au.

 In particular, \cite{Lykawka2007} found that evolution of scattered TNOs is described by multiple temporary resonance sticking and continuous scattering by Neptune and that this mechanism is relevant mostly at $a < 250$ au, consistent with our result. \cite{Yu2018} predicted that the current transient-sticking population comprises $40\%$ of the total transiently stuck + scattering TNOs, and therefore is a very important mechanism in the SD zone.

However, the general path through this region is a ``hand off'' from the gravitational control of one planet to another as was already found by Levison and Duncan (1997). Once a particle reaches a region near Jupiter and experiences close encounters with this planet, it rapidly suffers either an ejection or an injection into the inner region of the Solar System. This is connected to the correlations of mean lifetime with perihelion distance already mentioned.

\cite{Bailey2009} found that the long-term orbital evolution of Centaurs exhibit two types of behavior that are strongly correlated with Centaur lifetime. Centaurs with shorter lifetimes are characterized by diffusive evolution of semimajor axis and the ones with longer dynamical lifetimes  are dominated by resonance hopping. Those mechanisms were also reported by \cite{Tiscareno2003} and our previous work DB07.

\section{GPC and Centaurs from other sources}
\label{otrasfuentes}
The zone of $q < 30 $ au is also fed by other minor body populations. Here we review the works about other contributions to that zone and compare them with the main contribution from the SD.

\subsection{From Plutinos and other TN MMR}
\label{plutinosyres}

The $2 : 3$ MMR with Neptune, i.e. the Plutinos is the most populated observed resonance in the TN region. \citet{Morbidelli1997} analyzed the dynamical structure of the $2 : 3$ MMR and found that there exists a slow chaotic diffusion zone that should be an active source of Neptune-encountering  bodies at current epoch of the Solar System and then a source of current Centaurs and JFCs. However, they found that only 10 $\%$ of the Plutinos in this weakly chaotic zone have been delivered to Neptune-encountering orbits in the last Gyr. \cite{Tiscareno2009} performed numerical integrations over 1 Gyr timescale on objects in the $2 : 3$ and $1 : 2$ MMR with Neptune. They found that the escaped Plutinos and Twotinos spend roughly equal amounts of time as Centaurs as they do as SDOs. However, only $ \sim 20 \%$ of both resonant objects survived the 1-Gyr integration, and only $\sim 15 \%$ are projected to survive for 4 Gyr. Thus, although the rate of escape should be high at initial times, it would be low during the present epoch in agreement with \citet{Morbidelli1997}. 

The collisional evolution of resonant objects could be another way of removal from the resonance; however, \cite{deelia2008} found a flux rate of escape of $0.5 \%$ of Plutinos in 10 Gyr, which is much less than the dynamical removal. \citet{DiSisto2010} analyzed the contribution of Plutinos to the GPC zone and found that almost all the Plutinos that escape from the resonance enter the GPC zone with a current rate of injection of $1.62 \times 10^{-10} yr^{-1}$, i.e., $\sim 3$ times less than the injection from the SD found by DB07. They considered the number and size distribution of Plutinos proposed by  \cite{deelia2008} which is given by:
\begin{xalignat}{4}
	N_P(>D) &= C \bigg(\frac{1 \text{km}}{D}\bigg)^{p} &&\text{for} && D \leq 60~\text{km},  \nonumber \\
	N_P(>D) &= \text{7.8} \times 10^{9} \bigg(\frac{1 \text{km}}{D}\bigg)^{3} &&\text{for} && D > 60~\text{km},
	\label{nplutinos}
\end{xalignat} 
where $C = \text{7.9} \times 10^{9} \times 60^{p-3}$ by continuity for $D$ $=$ 60 km and two values are considered for the cumulative power-law index $p$: 2.5 and 1.5.

\citet{Alexandersen2016} performed a detailed study and analysis of the Plutino size distribution and found the cumulative number of Plutinos to be $N_P(H < 8.66)  = 9000 \pm 3000$ and $N_P(H < 10)  = 35000 \pm 10000$ while from Eq. (\ref{nplutinos}), considering an albedo of 0.06, the corresponding numbers in terms of diameter are $N_P(D> 100$km) $= 7900$, and $N_P(D> 60$km) $= 36500$, both in perfect agreement with \cite{Alexandersen2016}.  
Therefore, \citet{DiSisto2010} calculated from the Plutino SFD and the rate of injection to GPC that there could be between $1.8 \times 10^{6} $ and $1.8 \times 10^{7}$ GPC with $D > 1$ km from Plutinos in the current population. This is an order of magnitude less than that from the SD. This contribution together with that of the SD and from Jupiter Trojans can be seen in Fig. \ref{ntgpc}.
Using the simulation from \citet{DiSisto2010} it is possible to compute the rate of injection of Plutinos to the Centaur zone ($a < 30$ au). We find that $80 \%$ of the Plutinos that escape from the resonance reach the Centaur zone. Extracting from the output orbit files the first time a Plutino enters this zone, we obtain that their rate of injection is of $Y_p  = 1.316 \pm 0.002 \times 10 ^{-10} N_{P}/year$. \cite{Munoz2019} obtained a rate of injection from the TN resonant population to the JFCs zone (defined by $2 < T < 3$ and $q < 2.5$ au). They found that the resonant regions that contribute mainly to JFCs are the 3:2 and 5:2 MMRs and the total rate of injection to JFCs is of  $Y_p  = 1.07  \times 10 ^{-10} N/year$ by including as massive particles 34 large TNOs and  $Y_p  = 0.954  \times 10 ^{-10} N/year$ without those large TNOs. Both \cite{Munoz2019} rates are very similar and slightly smaller than our rate since they are calculated for injection to JFCs an our calculation is for Centaurs.
We also calculate the Plutino mean lifetime as a Centaur to be $l_{PC}=8.8$ Myr. Therefore, proceeding in the same way as for SDOs (see Eq. (\ref{npc})),  but considering the injection rate and mean lifetime for Plutinos to Centaurs and the number of Plutinos from Eq. (\ref{nplutinos}), we calculate the cumulative number of Plutinos in the Centaur population. 
 This contribution together with that of the SD and from Jupiter Trojans can be seen in Fig. \ref{ntc}.

\citet{DiSisto2010} also studied the dynamical evolution of escaped Plutinos in the GPC zone and found that they behave according to the four dynamical classes proposed by DB07. However, the resonance hopping mechanism and a high frequency of Kozai resonances in all the four classes are preferred. In particular, MMRs and Kozai resonances in the zone of $30 < a < 50$ au are more frequent than others. They also noted that some escaped Plutinos are captured again on the 3:2 MMR for some time before reaching their final state. The dynamical evolution of escaped Plutinos is very similar to that of SDOs in the GPC zone as can be seen by comparing the time-weighted distribution maps (see Fig. 3 in \cite{Disisto09}) with our Fig. \ref{mapaae} and \ref{mapaai}. 

\cite{Horner2010} evaluated the possibility of Neptune Trojans being a source of GPC. They found that 1 km-sized Neptune Trojan enters the region of $q < 30$ au every $\sim  60 - 200$ years, a rate well below the one from the SD of 3 km-sized SDOs every 2 years.
 \cite{Volk2013tesis} calculated the supply rate of GPC and JFCs from each of the sub-populations in the TN region with numerical simulations, obtaining a fractional escape rate of $3 \times 10^{-11} yr^{-1}$ for the $3:2$ MMR and  $ 10^{-10} yr^{-1}$  for the $1:2$ MMR and similar for the $5:3$ and $7:4$, well below the escape rate from the SD.
\citet{Alexandersen2016} estimate also populations of Neptunian and Uranian coorbitals, as well as objects in the $3:1$ and $4:1$ MMR with Neptune which are two orders of magnitude less than the Plutino population. Therefore, their contribution to the GPC and Centaur zones would be negligible.

 However, with the recent surveys on the TN zone, several MMRs with Neptune have been discovered to be populated.  \cite{Gladman2012} discuss the MMRs in the TN zone using objects detected by CFEPS, and provide absolute population estimates for those resonances. They found that the 5:2 MMR could be as populated as the 3:2 whereas the 2:1 MMR has $\sim 4$ times less objects than the 3:2 and 5:2 MMRs. For the other MMR: 4:3, 5:3, 7:3, 5:4, 7:4, 3:1, and 5:1, they measure significant populations being the 5:1 $60 \%$ of plutinos. This last MMR was also found to be one of the most populous resonances by \cite{Pike2015}. By analyzing the resonant objects discovered by OSSOS survey, \cite{Volk2016} found that the 2:1 MMR could be as populated as the 5:2 MMR and possibly as numerous as plutinos. Therefore, the real contribution of other MMRs than 3:2 could be as important as plutinos, but more work is needed in relation to their dynamical evolution to Centaurs and JFCs.

\subsection{From CTNOs}
\label{ctnos}
Classical TNOs and Plutinos were the first to be discovered among TNOs, due to their relatively closeness and cold orbits. A few years after the first classical object was discovered in 1992 by David Jewitt \citep{Jewitt1992},  \citet{Levison1997b} performed the first dynamical evolution of CTNOs.  They  selected from a previous simulation, 20 particles that left the Kuiper Belt (KB) after being stable for over 1 Gyr. Thus, the orbits of these particles were representative of the orbits of objects currently leaving the classical TN region. They added clones of those particles and followed their dynamical evolution for 1 Gyr under the gravitational action of the Sun and the four Giant planets up to collision or ejection. They investigated the distribution and fates of objects which had left the KB in the current configuration of the Solar System. They found that the objects evolve inward from the KB being under the dynamical control of one planet at a time and keeping a very narrow range in the Tisserand parameter (T) with respect to each control giant planet. Thus, they reach the Jupiter control with T just below 3 in a very narrow range as it is observed in JFCs. They also estimated a number of km-sized ecliptic comets of $\sim 1.2 \times 10^7$. 

Based on the study of the TN region as a source of JFCs and Centaurs,  \cite{Volk2013tesis} obtained a  fractional escape rate of $ 0.55 \times 10^{-10} N/yr$  for the debiased CTNOs, a rate that is lower than our results for the escape rate from the SD. In fact, \cite{Volk2013tesis} found that the contribution to Centaurs from the SD is at least two times greater than that of CTNOs.  
\cite{Munoz2019} obtained a rate of injection from the CTNOs to the JFCs zone of  $0.306  \times 10 ^{-10} N/year$ with the integration including 34 large TNOs and  $ 0.261  \times 10 ^{-10} N/year$ without the large TNOs. Those rates are slightly smaller than the ones obtained by \cite{Volk2013tesis} since they calculated the escape rate and \cite{Munoz2019} calculated the injection to JFCs.

\subsection{From Jupiter Trojan asteroids}
\label{ast}

Jupiter Trojans are asteroids trapped in the $1:1$ MMR with Jupiter and located around the $L_4$ and $L_5$ Lagrangian points on relatively stable orbits. However, they cover a space near $L_4$ and $L_5$ where zones of different scales of stability can be differentiated. This allows some Trojans to escape from the resonance on timescales that are shorter than the age of the Solar System \citep{Levison1997a,Disisto2014}. Those escapees may contribute to populate other zones of the Solar System.
The contribution of Jupiter Trojans to other Solar System populations was studied by \cite{Disisto2019}. They considered the observed Jupiter Trojan population and performed long-term numerical simulations to study the trajectories of those Trojans that leave the resonance. They obtained a constant escape rate from both Lagrangian points over time of $Y_{L_4} = 7.0398 \times 10^{-11} \pm 8 \times 10^{-14} N_{T}/ yr$ and $Y_{L_5} = 7.5590 \times 10^{-11} \pm 13 \times 10^{-14} N_{T}/ yr$,  where $N_T$ is the cumulative number of Trojans. 
\cite{Disisto2019} also found that  $\sim 90 \%$ of escaped Trojans from $L_4$ and $L_5$ go through the GPC and Centaur zone  and that the distribution of both spatial and angular orbital elements of escaped Trojans are similar to those of the observed Centaurs (see their Fig. 10). They obtained the number of Trojans ejected out of the resonance per year as a result of dynamical evolution considering the number of Trojans and their size distribution and the member number asymmetry between the $L_4$ and $L_5$ swarms. Additionally, including the rate of escape from the resonance by collisional evolution from \cite{deelia2007}, they calculated the number of escaped Trojans in each minor body population and found that the contribution of escaped Trojans in the Centaur zone would be minor.

The contribution from both Jupiter Trojan swarms to GPC and Centaurs are plotted together with Plutinos and SDOs in Fig. \ref{ntgpc} and \ref{ntc}.

\subsection{Retrograde GPC and Centaurs}
\label{retro}

Other sources of Centaurs apart from TNOs have been proposed and studied, especially in order to explain particular cases. There is a very low probability for a high inclination Centaur or even a retrograde one to have its origin in the TN region \citep{Volk2013}.  \cite{Brasser2012} showed that Centaurs with inclinations $i > 70^{\circ}$  and $q >15$ au,  originate mainly in  the Oort Cloud being decoupled from it by the gravitational perturbations of Uranus and Neptune alone. Those authors estimated that  there are between 1 and 200 Centaurs with absolute magnitude $H < 8$ ($D  \gtrsim  150$ km for $p_v = 0.05$) in that range of $i$ and $q$. The case of retrograde Centaurs was analyzed by \cite{delafuente2014} who concluded that they formed an heterogeneous group that may have its origin in the Oort cloud but that other sources or dynamical mechanisms that enlarge the inclinations can not be ruled out. Those Centaurs are trapped in transient retrograde, mostly higher-order resonances with the giant planets, leading to chaotic diffusion and  thus making their orbits dynamically unstable.
This is also observed for the retrograde GPC in our dynamical evolution of observed objects in Sect. \ref{evolution}. 
\cite{Nesvorny2017} performed numerical simulations of the early evolution of the Solar System in which cometary reservoirs were formed and evolved, and found that the SD (with a flat inclination distribution) is the source of  ecliptic comets, while the Oort cloud is the source of Halley-type comets (HTCs) with an isotropic inclination distribution.
\cite{Nesvorny2019} performed a study directed to test a model of the early evolution of the outer Solar System by running it through the OSSOS survey simulator. They started their simulation with an original flat planetesimal disk below 30 au, and followed its evolution from the onset of Neptune's migration to the present time. They obtained that in the last Gyr, $11 \%$ of Centaurs evolved from the region with $a > 5000$ au. However, only some very-high-inclination Centaurs can be explained by their simulation and then other sources of very-high-inclination and retrograde Centaurs are needed.

\subsection{Region interior to Saturn and Jupiter crossers}
\label{jup}

The region between Jupiter and Saturn has been investigated by a number of authors, analyzing both the physical and dynamical processes of its objects. However, the Jupiter crossers zone is difficult to address completely. The numerical simulations that study the TNOs as a source of GPC or Centaurs generally consider only the perturbations of the Giant planets; they discard the particles when they reach Jupiter's orbit since the terrestrial planets and a lower integration step is necessary for accurate results on dynamical evolution.  On the other hand, a purely dynamical simulation would be no longer appropriate in the region interior to Jupiter's orbit (the JFC zone) because of erosion processes of icy objects at this distance.  Therefore, a general self-consistent physico-dynamical model from the TN zone to the JFC zone is difficult to address. This problem was approached by \cite{Disisto09} who continued the integration of the particles that arrived at Jupiter in the simulation of DB07. In this work, \cite{Disisto09} integrated those particles plus clones under the perturbation of all the planets and by considering also non gravitational forces and a model for sublimation and splitting of comets. They focused their study in JFCs ($P < 20$ yrs and $2 < T < 3.1$) by obtaining their distribution  in the orbital element space and also their expected number in regions of perihelion distances. They obtained a mean lifetime of JFCs with $D > 2$ km and $q < 1.5$ au of about 150–200 revolutions ($\sim 1000 $ yrs), for $q < 2.5$ au $\sim 300 - 450$ revolutions and a total population of JFCs with $D > 2$ km  within Jupiter's zone of $450 \pm 50$. However, they found a greater  population of non-JFCs (those that don't fulfill the conditions $T > 2$ and $P < 20 $ yr simultaneously), which would be 4 times greater for Jupiter-crossing orbits of comets of $D > 2$ km. Thus leading to a whole population of JFCs + non-JFCs of  $D > 2$ km of $2250 \pm 250$ in the zone of $q < 5.2$. \cite{Nesvorny2017} performed numerical simulations of a primordial Solar System, in which cometary reservoirs are formed and evolved over 4.5 Gyr. From this simulation they found the current population of JFCs and also HTCs. By accounting for the physical lifetime of active comets through different parametrizations, and comparing with observations, they inferred  a mean lifetime of JFCs with $q < 2.5$ au of $\sim 300 - 800$ revolutions, consistent with \cite{Disisto09}. They also found that the number of JFCs with $D > 10$ km is $\sim 1 - 2$.

\cite{Disisto09} also found that larger comets usually return to the Centaur zone in their evolution, but smaller ones ($1-$ km comets), suffer a quick erosion reaching a minimum proposed radius ($100$ mts) before they can evolve into other dynamical states. Therefore, the Centaur zone near Jupiter is partially re-filled by comets and this is strongly dependent on the size. Working with the numerical simulations of \cite{Disisto09}, we found that objects that return to the Centaur zone from the JFC zone experience a very quick dynamical evolution among the Giant Planets until ejection. Their dynamical mean lifetime in the region of $q > 5.2$ au  and $a < 30$ au is of $0.4$ Myr, very similar to that of escaped Jupiter Trojans. Also their dynamical evolution is similar to that of JTs (see Fig. 3 in \cite{Disisto2019}).

\cite{Brasser2015} performed numerical simulations of the evolution of SDOs until they became visible JFCs, keeping track of their number of perihelion passages with  $q < 2.5$ au. They used a simple fading law applied to JFCs that depended on the number of revolutions.  From their simulation and observational data on JFCs, they estimate that the steady-state number of active JFCs with $q < 2.5$ au and diameter $D > 2.3$ km
is $294^{+556}_{-235}$. This is a larger value than estimates by \cite{Levison1997a} and \cite{Disisto09}, but given the large uncertainty, it can be considered to be in the same range as the latter. 

On the other hand, in the Jupiter-Saturn region there are also some signs of sublimation in some objects in contrast to others that remain inactive. \cite{Fernandez2018} addressed a sub-population of Centaurs in the Jupiter-Saturn region,   with the aim of investigating the dynamical evolution and end states of active and inactive Centaurs. The authors performed the study with numerical simulations integrating two samples of real Centaurs (one for active Centaurs and the other for inactive ones) and clones under the gravitational action of the Sun and the planets from Venus to Neptune. They concluded that inactive and active Centaurs have different dynamical behaviors and different median dynamical lifetimes that could explain the presence of activity in some Centaurs and the lack of it in others. In addition, they show that probably evolution (such as a recent drastic drop in perihelion distance) and not intrinsic physical differences explains the display or not of activity of Centaurs in the Jupiter-Saturn region. 

Close encounters with Jupiter or Saturn could turn out to be drastic for a comet-like object. In a close encounter of a comet with one of these giant planets, tidal effects can become so important that they can cause the fragmentation of the body and its eventual collision with the planet, as was the case of Shoemaker Levy 9 with Jupiter in 1992-1994. 
Also, they could be important on the preservation of the rings around Centaurs. In 2014, \citet{Braga2014} discovered around Chariklo, the first ring around a Centaur. It was an unexpected discovery, since given the chaotic dynamical evolution of Centaurs, with close encounters with the giant planets, it should be difficult to retain a ring. Of course this depends on the competition between the formation processes (probably by collisions) and destruction processes. 
Numerical studies by \cite{Araujo2016} and \cite{Wood2017} evaluated the encounters of Chariklo (and clones) within 1 Hill radius of the planet and found that they have little effect on the rings and that they could survive through its evolution as a Centaur. However, \cite{Wood2018} found that a close encounter has a noticeable effect on the ring of a small body if it occurs at a distance within $\approx$ 1.8 tidal disruption distances (the distance where tidal forces can disrupt a small body–ring particle), which is much smaller that the Hill radius. Therefore, the presence and survival of rings on Centaurs is strongly dependent on their dynamical evolution in the giant planet zone.

Therefore, the region interior to Saturn up to Jupiter, is a complex zone to evaluate the dynamical behavior of Centaurs. Physical models are needed together with dynamical models, or at least it is necessary to consider the possible shortening of the physical lifetimes of small objects by the possible presence of activity. However, the dynamical evolution is fast.  
In the region of low-eccentricity orbits with $q > 5.4$ au and $Q < 7.8$ au, \cite{Sarid2019} found a niche which they called a Gateway for transition between Centaurs and JFCs. They found that  $77 \%$  of objects in this Gateway region will become or have already been JFCs, and nearly half of all JFCs will pass through this Gateway region before experiencing significant sublimation. Therefore, this is an important discovery in relation to the dynamical link between populations, but also for the purposes of the conditions to develop comet activity. 

\section{General model from all GPC and Centaur sources}
\label{todo}

\begin{figure}[h!]
\includegraphics [width=0.95\textwidth]{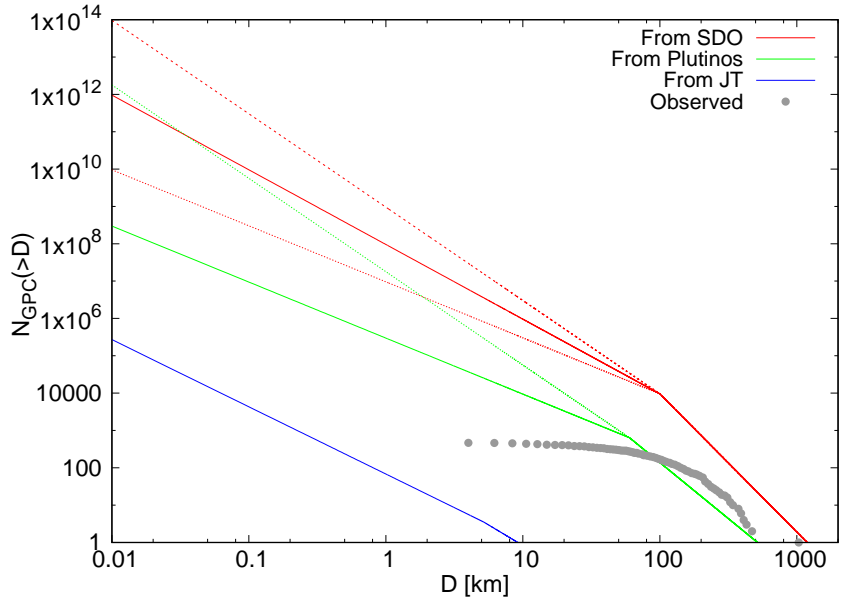}
\caption{ Cumulative number of Giant Planet Crossers ($N_{GPC}(>D)$) vs diameter $D$ from SDOs (red curves), Plutinos (green curves) and Jupiter Trojans (blue curves). The gray dots correspond to the observed population. For the observed population we use an albedo $p_v =0.06$ to convert absolute magnitude to diameter.}
\label{ntgpc}
\end{figure}

\begin{figure}[h!]
\includegraphics [width=0.95\textwidth]{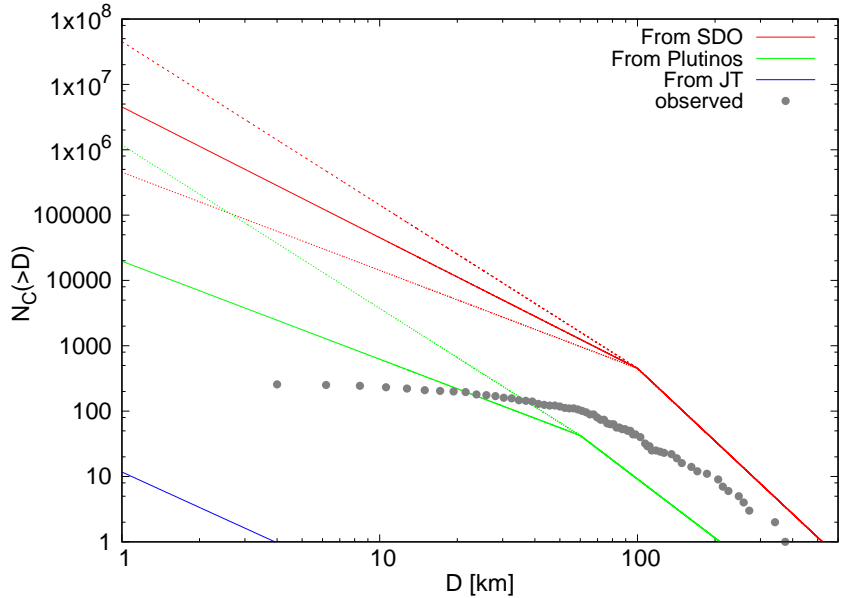}
\caption{ Cumulative number of Centaurs ($N_{C}(>D)$) vs diameter $D$ from SDOs (red curves), Plutinos (green curves) and Jupiter Trojans (blue curves). The gray dots correspond to the observed population. For the observed population we use an albedo $p_v =0.06$ to convert absolute magnitude to diameter.}
\label{ntc}
\end{figure}

From the results obtained in the previous sections, we show the combined  contribution of SDOs, Plutinos and Jupiter Trojans to GPC and Centaurs. In Figs. \ref{ntgpc} and \ref{ntc} the cumulative number of GPC and Centaurs respectively are plotted against diameter.  
As can be seen from those figures, the GPC and Centaur zones are mainly fed by SDOs from the TN zone, and the other contributions are secondary. It is notable that both the intrinsic number of GPC and Centaurs proposed by our model are respectively greater than the real observed numbers. However, the difference between model and observation for Centaurs is smaller than for GPCs. This is due to biases in the discovered populations which tend to favor the detection of Centaurs over GPCs, since the latter are more distant objects. Hence, there could be a great number of GPCs to be discovered yet.  

Those results are very sensitive to the size distribution of the source populations, and this is not a concluded topic. Especially for small diameters, where the SFD of objects may have a break, the results are very variable. We have shown in  Fig. \ref{ntgpc} and \ref{ntc} limit values for the power law index of the size distributions at smaller sizes.  This uncertainty leads to, for example, that for $D <5 $ km, depending on the size distribution index, the contribution from Plutinos could be greater than the contribution from SDOs. Although the SDOs and Plutinos are two dynamically excited populations and could probably have the same initial index in their size distribution (or at least similar), the collisional evolution of Plutinos could have changed this. Moreover, for very small objects ($\lesssim 1$ km), there may be an additional break in the SFD slopes of TNOs as was suggested from cratering counts \citep{DiSisto2013, Singer19}, but more studies are needed for those small TNOs.
Therefore, this is an open scenario, and more observations and studies are needed to assess completely the size distribution at small diameters. 

On the one hand, secondary sources may explain some peculiarities, probably related in particular to different Centaur compositions. Although JT are a minor source of GPC and Centaurs, \cite{Disisto2019} found that some NEO-JFC orbits, Encke-type comets and impacts on Jupiter as the Shoemaker Levy 9 case can be explained by the dynamical evolution of escaped Trojans. 
On the other hand, in their path to being ejected from the Solar System, JT and JFCs go faster through the giant planet zone than in their way inward from the SD and other secondary sources. They are also affected by the physical processes they suffered when they were within the orbit of Jupiter, so they could give rise to different types of Centaurs mainly in the area near Jupiter and Saturn.

\section{Conclusions}
\label{conclusiones}

For the present review we have performed new numerical simulations and calculations of the Giant Planetary Crossers studying their number and their evolution from their sources, considering the current configuration of the Solar System.  

From the dynamical evolution of SDOs, we have calculated the number of Centaurs predicted by our model. In comparison, the recent estimates by OSSOS team are somewhat lower than ours for the diameters studied (see Fig. \ref{ncent}).

From the results obtained in the numerical simulation of SDOs and from previous works, we obtain the contribution of SDOs, Plutinos and Jupiter Trojans (JT) to GPC and Centaurs as shown in Figs. \ref{ntgpc} and \ref{ntc}. From those Figs., the SD is the main source of prograde GPC and Centaurs as was first suggested by \citet{Duncan1997} and found also by other subsequent studies as DB07 and \cite{Volk2013tesis}. For example, there are 9600 GPCs from the SD with $D > 100$ km and $ 10^8$ with $D > 1$km (with a differential SFD index of 3). 

The cumulative number of escaped Plutinos in the GPC zone lies between one and two orders of magnitude less than that from the SD. However, those results are very sensitive to the size distribution of the source populations and especially for small diameters, where the SFD of objects may have a break. The uncertainty at small sizes could lead to an important contribution from Plutinos if the index of the SFD of Plutinos was greater than that of the SDOs. New observations and works on other MMRs in the TN zone \citep{Gladman2012, Pike2015, Volk2016} indicate that some MMR, specially the 2:1 and 5:2, could be as populated as plutinos,  and therefore they have to be in consideration as possible contributors to Centaurs and JFCs.
The contribution from JT can be considered negligible although it may explain some peculiarities. 

The observed number of GPCs and Centaurs are well below our model due to observational biases in the surveys. Therefore, there could be many GPCs and Centaurs to be discovered yet.  

The dynamical evolution of TNOs in general through the GPC zone and then to the JFC region, can be seen as a “hand off” from the gravitational control of one planet to another as was first shown by \cite{Levison1997b}. The mean lifetime of SDOs in the GPC zone is of 68 Myr. There is a correlation between the particles lifetime in the GPC zone and their perihelion distance, where the greater the perihelion, the longer the mean lifetime. This is connected with a slower evolution of objects at larger perihelion, where the variation of $q$ is small in the region between the orbits of Saturn and Neptune. 
We detect transfers between mean motion resonances, known as resonance sticking, in the zone of $\sim 30$ au $ < a \lesssim 200$ au. Beyond 200 au, the general dynamical evolution is a low increase of semimajor axis  but keeping the perihelion distances near Neptune. Thus, the objects are continuously entering and leaving the GPC zone up to ejection (injection to the GPC zone of those particles is more rare).

The initial inclination affects the mean lifetime of particles as GPC, being the ones with high initial inclinations the ones that survived longer time in the GPC zone. In the Jupiter-Saturn region, the dynamical evolution is faster. Different dynamical behaviors were also found by \cite{Fernandez2018} in inactive and active Centaurs, that are also connected with different median dynamical lifetimes and that may help to explain the activity or lack of activity in some Centaurs. 

From our simulations, the initial particles (i.e. objects in the SD) which had initial inclinations lower than $55^{\circ}$, evolved into prograde GPC and Centaurs. Thus, with our model we were not able to generate any prograde GPC with inclinations greater than $60^{\circ}$ nor a retrograde GPC. Therefore, as was found by other authors, retrograde Centaurs probably did not come from the SD but from other source such as the Oort Cloud (e.g. \cite{Brasser2012}, \cite{Volk2013}). 
In fact, we have performed numerical simulations of the dynamical evolution of observed GPC, within which there are $17$ retrograde objects that show different properties than the prograde population. In this study, the observed prograde population has a similar behavior to those GPCs that enter from the SD (from our model), but retrograde GPCs have a comparatively shorter median lifetime. Thus, retrograde GPCs experience a faster evolution than prograde ones. 
 However, it  is  probable  that this faster evolution is due to the fact that the majority of retrograde GPCs (15 of the 17) have low perihelion values and then, lower lifetimes. In fact the two retrograde objects with high perihelion distances have long lifetimes, greater than 600 Myr. 
We also note that some retrograde objects experience MMR captures as well. 

\section{Acknowledgments}
\label{agradecimientos}
We acknowledge the financial support given by IALP, CONICET and of Agencia de Promoción Cient\'{\i}fica and also the FCAGLP for extensive use of their computing facilities. We would like to thank two anonymous reviewers for their detailed comments and suggestions which helped us to significantly improve this article, and also Tabar\'e Gallardo for valuable discussion about encounter conditions and Opik encounter probability.

\bibliography{biblio.bib}  
\end{document}